\documentclass[
 10pt,
 twocolumn,
 superscriptaddress,
 nofootinbib,
 amsmath,amssymb,
 acmart,
 pra,
]{revtex4-1}
\usepackage[utf8]{inputenc}
\usepackage{braket}
\usepackage{hyperref}
\usepackage{graphicx}
\usepackage{tikz}
\usepackage{xcolor}
\usepackage{mathtools}
\usepackage{soul}
\setstcolor{blue}
\usepackage{comment}
\usepackage{listings}

\newif\ifdraft
\drafttrue
\ifdraft
\newcommand{\note}[1]{ {\textcolor{orange} { **: #1 }}}
\newcommand{\alnote}[1]{ {\textcolor{red} { ***Andre: #1 }}}
\newcommand{\crnote}[1]{ {\textcolor{teal} { ***Carlos: #1 }}}
\newcommand{\jknote}[1]{ {\textcolor{blue} { ***Johannes: #1 }}}
\newcommand{\smnote}[1]{ {\textcolor{purple} { ***Shaoming Eh: #1 }}}
\else
\newcommand{\note}[1]{}
\newcommand{\alnote}[1]{}
\newcommand{\crnote}[1]{}
\newcommand{\jknote}[1]{}
\newcommand{\smnote}[1]{}
\fi

\begin{document}

\title{A practical overview of image classification with variational tensor-network quantum circuits}

\author{Diego Guala}
\affiliation{Xanadu, Toronto, ON, M5G 2C8, Canada}
\author{Shaoming Zhang}
\affiliation{BMW AG, Munich, Germany}
\affiliation{Department of Informatics, Technical University of Munich}
\author{Esther Cruz}
\affiliation{Max-Planck-Institute of Quantum Optics, Hans-Kopfermann-Straße 1, 85748 Garching, Germany}
\author{Carlos A. Riofr\'io}
\affiliation{BMW AG, Munich, Germany}
\author{Johannes Klepsch}
\affiliation{BMW AG, Munich, Germany}
\author{Juan Miguel Arrazola}
\affiliation{Xanadu, Toronto, ON, M5G 2C8, Canada}
\date{\today}

\begin{abstract}
    
    Circuit design for quantum machine learning remains a formidable challenge. Inspired by the applications of tensor networks across different fields and their novel presence in the classical machine learning context, one proposed method to design variational circuits is to base the circuit architecture on tensor networks. Here, we comprehensively describe tensor-network quantum circuits and how to implement them in simulations. This includes leveraging circuit cutting, a technique used to evaluate circuits with more qubits than those available on current quantum devices.
    We then illustrate the computational requirements and possible applications by simulating various tensor-network quantum circuits with PennyLane, an open-source python library for differential programming of quantum computers.
    Finally, we demonstrate how to apply these circuits to increasingly complex image processing tasks, completing this overview of a flexible method to design circuits that can be applied to industrially-relevant machine learning tasks.
\end{abstract}

\maketitle

\section{Introduction}
Tensor networks have been studied for decades across several disciplines, most notably in the context of many-body quantum systems~\cite{RevModPhys.93.045003,  TN-Intro}. 
In quantum computing, tensor networks have been used for the classical simulation of quantum computers~\cite{markov2008simulating,biamonte2017qml,huang2020classical, pan2021solving} and as a framework to build new machine learning models~\cite{stoudenmire2017supervised, martyn2020entanglement, han2018unsupervised}. Such studies have sparked interest in understanding whether tensor networks can be applied to inspire circuit design in the field of variational quantum algorithms~\cite{cerezo2021variational, Schuld2014, biamonte2017qml}.\\

Pioneering work combining tensor-network architectures and variational quantum algorithms was reported in Refs.~\cite{Huggins2019, cong2019quantum, haghshenas2021optimization, haghshenas2022variational}. The main idea is to design quantum circuits replicating tensor network architectures such as tree tensor networks and matrix product states~\cite{TN-Intro,murg2015tree}. We refer to the resulting circuits as \emph{tensor-network quantum circuits}. These quantum circuits may also provide a way to process tensor networks that are too large for classical methods~\cite{foss2021entanglement}.\\

As quantum computing technologies mature, they must become accessible to a broader community of scientists, engineers, and practitioners. This is fundamental for the development of the field, as accessibility leads to more ideas that can be tested and potentially commercialized~\cite{Bayerstadler2021}. To lower the barrier of entry for practitioners interested in studying quantum tensor-network methods, this work comprehensively describes tensor-network quantum circuits and how to implement them in practice. In Sec.~\ref{sec:priorwork}, we guide the reader through the process of generating quantum circuits from tensor network architectures. This includes an explanation of how to combine quantum circuit cutting techniques~\cite{peng2020simulating,tang2021cutqc,lowe2022fast,bravyi2016trading,piveteau2022circuit,perlin2021quantum,dunjko2018computational} with tensor-network quantum circuits, permitting more efficient classical simulation and providing a path for executing large circuits on devices with fewer qubits. We provide explicit formulas for the resource requirements to perform these cuts and discuss how the resources depend on the features of the original tensor network. In Sec.~\ref{sec:numdemo}, we apply the above framework to perform a variety of numerical experiments. This includes benchmarks for a suite of matrix product state (MPS) quantum circuits at various bond dimensions and qubit numbers, evaluated using circuit-cutting techniques. Finally, we demonstrate applications of tensor network quantum circuits by applying them to two tasks: binary classification of simple synthetic data and defect detection in welded-metal images.

\section{Variational Tensor-Network Quantum Circuits}\label{sec:priorwork}

\subsection{Tensor Networks}
The basic building blocks of tensor networks are tensors: multi-dimensional arrays of numbers~\cite{biamonte2017tensor}. Intuitively, tensors can be interpreted as a generalization of scalars, vectors, and matrices. Consider a two-dimensional array or matrix, $T$. The elements of this array can be indicated by $T_{ij}$,
where the index $i$ indicates the rows and the index $j$ indicates the columns of the matrix. Using the tensor network nomenclature, $T$ is a rank two tensor. A tensor's rank is the number of indices in the tensor---a scalar has rank zero, a vector has rank one, and a matrix has rank two. While the number of dimensions of an array is equivalent to the rank of the tensor, the length in each dimension is captured by the number of possible values an index can take. This is the index dimension.

\begin{figure}[t!]
    \centering
    \includegraphics[width=0.45\textwidth]{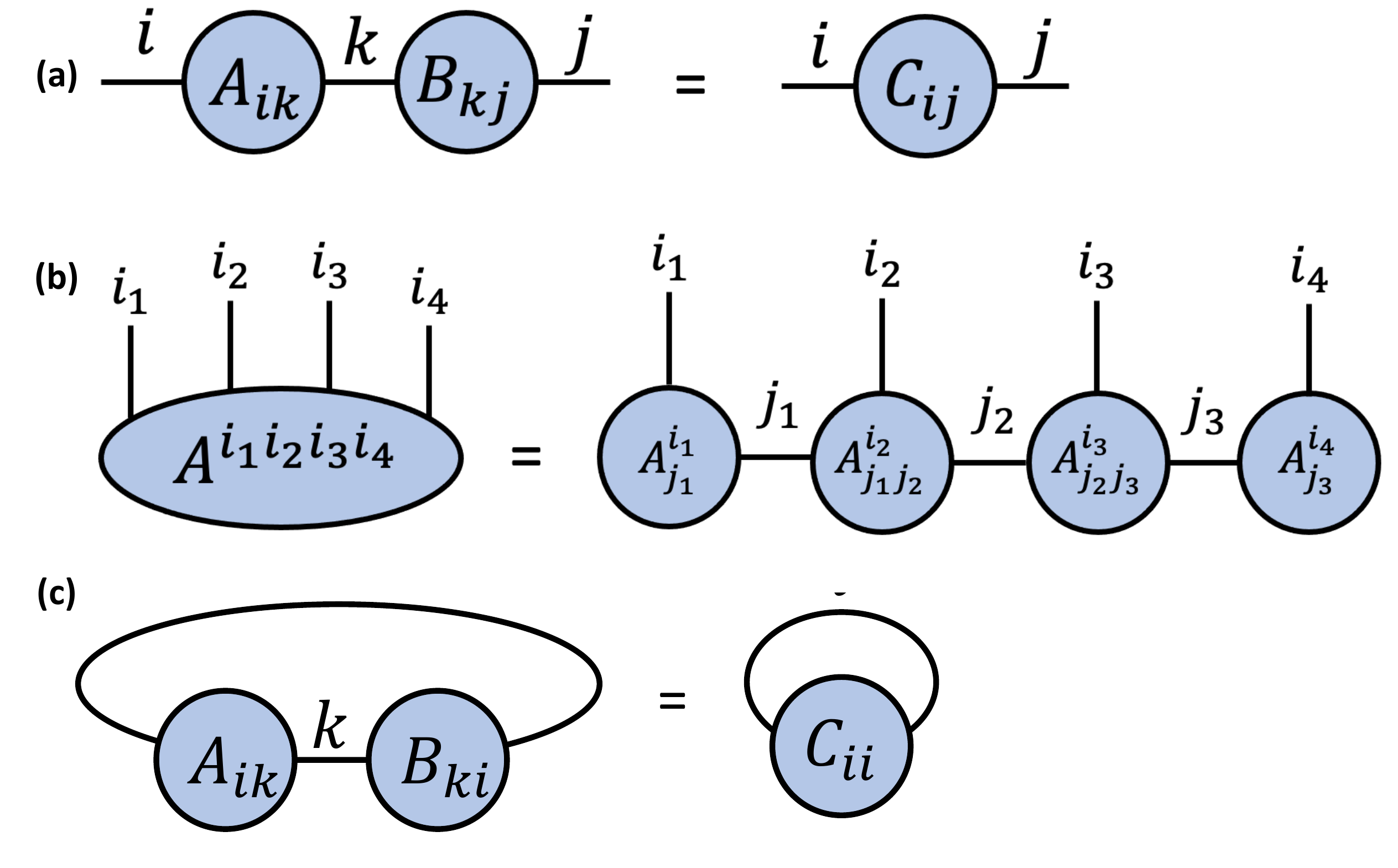}
    \caption{(a) A diagrammatic example illustrating the contraction of two rank-2 tensors (matrices) $A$ and $B$, connected by a repeated index $k$. Tensors can be contracted by summing over repeated indices. In this case, the contraction corresponds to the summation over $k$, as given in Eq.~\ref{eq:contraction}; (b) An example of a tensor $A^{i_1,i_2,i_3,i_4}$ and its factorization into matrix product state form $A^{i_1}_{j_1}A^{i_2}_{j_1j_2}A^{i_3}_{j_2j_3}A^{i_4}_{j_3}$, as given in Eq.~\ref{eq:MPS_A}. This factorization is done by singular value decomposition~\cite{SCHOLLWOCK201196}; (c) An example of the trace of tensors. The trace is equal to the connection of two different legs of the tensor and summing over the corresponding index.}
    \label{fig:tensors}
\end{figure}

A key operation in tensor networks is \emph{contraction}. Two tensors are contracted when they are combined into a single tensor by summing the product of  their respective entries over a repeated index.
For example, the standard matrix multiplication formula can be expressed as a tensor contraction
\begin{equation}
    C_{ij} = \sum_{k}A_{ik}B_{kj},
    \label{eq:contraction}
\end{equation}
where $C_{ij}$ denotes the entry for the $i$-th row and $j$-th column of the product $C=AB$. Graphically, this operation can be represented as depicted in Fig.~\ref{fig:tensors}a. For technical and historical reasons, tensors can also be defined using indices as superscripts, e.g., the notation $A_{i}^j$ can also denote a rank-2 tensor.\\ 

From tensors, we can create tensor networks. A tensor network is a collection of tensors where a subset of all indices is contracted. It is helpful to discuss tensor networks using diagrams similar to Fig.~\ref{fig:tensors}. In this language, tensors are represented by shapes such as circles or squares, and edges symbolizing the indices. The rule for contraction can be displayed in a tensor network diagram by connecting tensors with edges, where two connected tensors are contracted~\cite{biamonte2017tensor}. Tensor networks can represent complicated operations involving several tensors, many indices, and sophisticated contraction patterns. When multiple contractions happen in a tensor network, the corresponding summations can be performed in different orders. The sequence in which the contraction is carried out is known as the \emph{contraction path}. This is an important concept, as a suitable contraction path will decrease the computational complexity of tensor network contraction.\\

\begin{figure}[t!]
    \centering
    \includegraphics[width=0.45\textwidth]{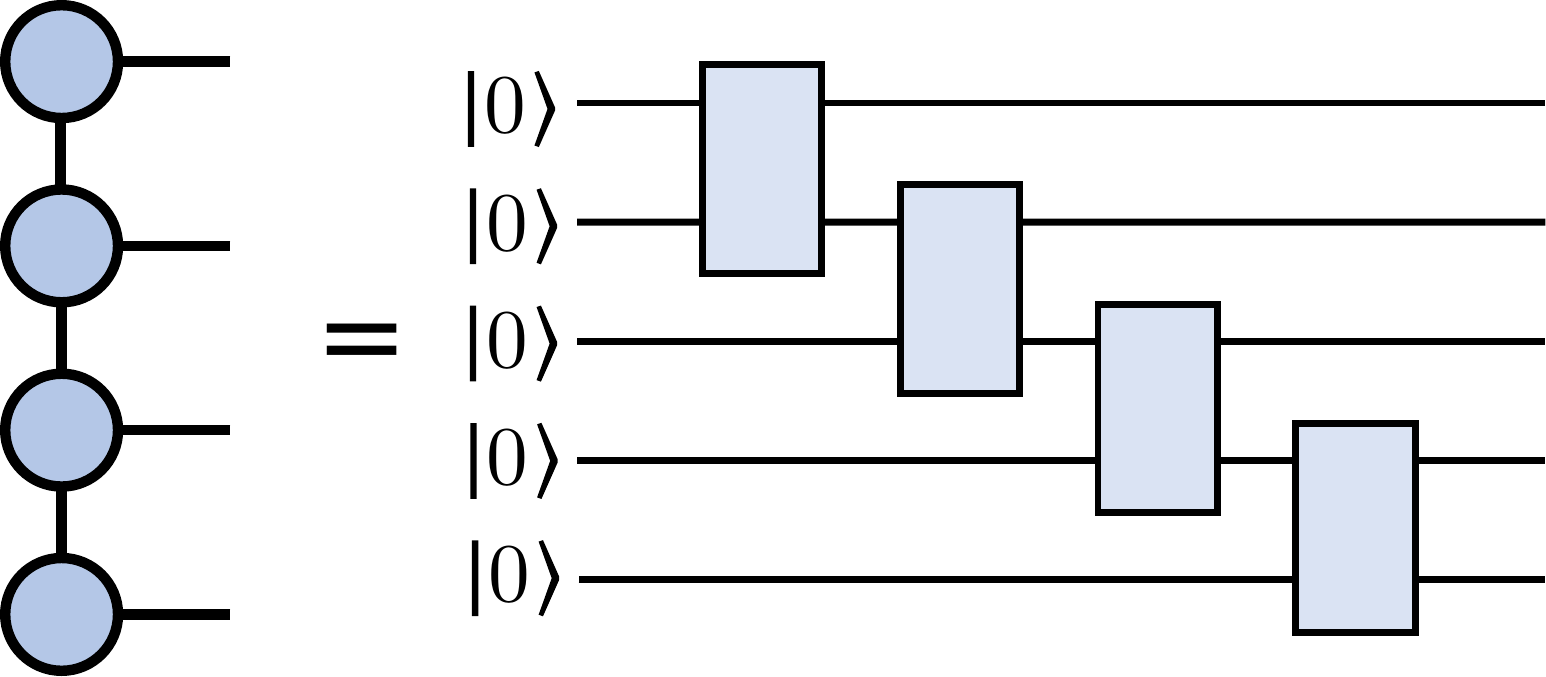}
    \caption{An example of a matrix product state (MPS) (\textit{left}) and the corresponding quantum circuit architecture (\textit{right}). Each tensor in the network is obtained by contracting the initial two-qubit state and a two-qubit gate.}
    \label{fig:MPS}
\end{figure}

Tensors of high rank can be difficult to work with since the number of array elements grows exponentially with the number of indices in the tensor.  A common strategy is to express high-rank tensors as a tensor network over tensors of smaller rank. For example, consider a tensor $A^{i_1, i_2, i_3, i_4}$ of rank four. It can be approximated by a tensor network of the form
\begin{equation}
    A^{i_1, i_2, i_3,i_4}\approx\sum_{j_1,j_2,j_3}A_{j_1}^{i_1}A_{j_1j_2}^{i_2}A_{j_2j_3}^{i_3}A_{j_3}^{i_4}.
    \label{eq:MPS_A}
\end{equation}
This tensor network is known as a matrix product state (MPS). It can be interpreted as a factorization of the tensor into a network consisting of tensors of smaller rank~\cite{TN-Intro}. An MPS factorization can be used to represent tensors exactly as long as the dimension of the internal $j$ indices, known as the \emph{bond dimension} $D$, is sufficiently large. Another option is to approximate tensors with matrix product states by selecting a smaller bond dimension, which can lead to simpler computations in exchange for lower accuracy. A graphical representation of an MPS is shown in Fig.~\ref{fig:MPS}b. Please see Refs.~\cite{biamonte2017tensor,TN-Intro} for more detailed introductions to tensor networks.

\subsection{Tensor-network quantum circuits}\label{sec:tnqc}
\begin{figure}[t!]
    \centering
    \includegraphics[trim={0 4cm 0 4cm},clip,width=0.45\textwidth]{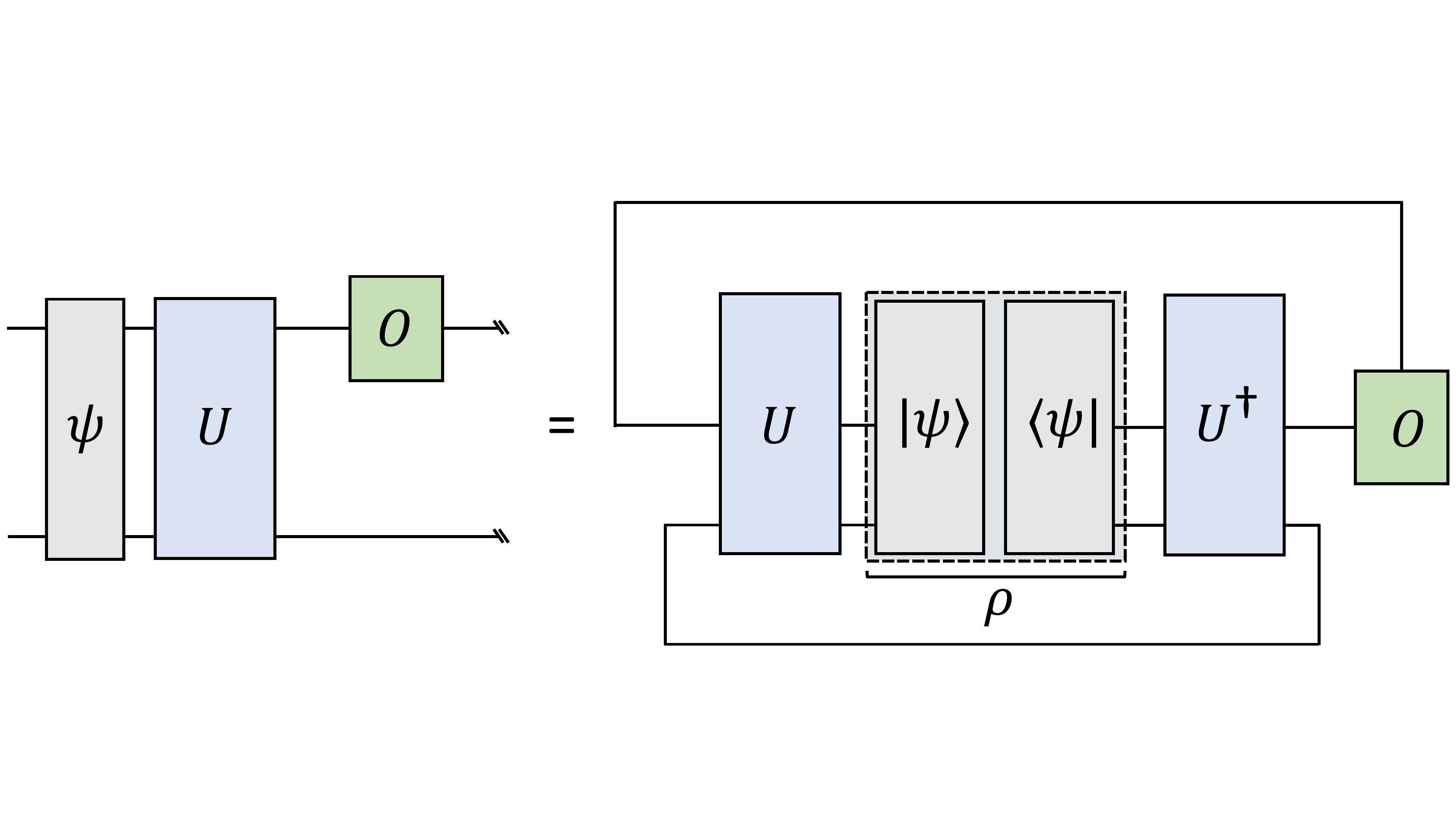}
    \caption{A quantum circuit and its corresponding tensor network. In the circuit-based picture, some initial quantum state $|\psi\rangle$ is evolved with a unitary gate $U$ and followed by a measurement of the observable $O$ on the top qubit (\textit{left}). This corresponds to the contraction of a tensor network and is similar to a trace operation, as given in Eq.~\ref{eq:mps} (\textit{right}). While other tensor network representations are possible, we choose a tensor network with bond dimension two, such that the wires in the circuit correspond to the tensor network legs.} 
    \label{fig:mpo}
\end{figure}

The connection between quantum computing and tensor networks can be seen by observing that quantum circuits can be expressed as tensor networks~\cite{peng2020simulating,shi2006classical,huang2020classical}. Formally, we consider a quantum algorithm consisting of an initial quantum state, represented by a density matrix $\rho$, a quantum circuit implementing a unitary transformation $U$, and a measurement performed on the output state $U\rho U^\dagger$ of the circuit, which is used to compute the expectation value $\text{Tr}(U\rho U^\dagger O)$ of an observable $O$. This expectation value can then be expressed by the tensor network
\begin{align}
    A(U)^{i_1i_2\ldots i_n}_{j_1j_2\ldots j_n} A(\rho)^{j_{1}j_{2}\ldots j_{n}}_{k_1k_2\ldots k_n} \,\, A(U^\dagger)^{k_{1}k_{2}\ldots k_{n}}_{l_{1}l_{2}\ldots l_{n}} \,\,A(O)^{l_{1}l_{2}\ldots
    l_{n}}_{i_1i_2\ldots i_n}\\=\text{Tr} (O U\rho U^{\dagger}),
    \label{eq:mps}
\end{align}
where sums are carried out implicitly over repeated indices, and $A(\rho)$, $A(U)$, $A(U^\dagger)$ and $A(O)$ are tensors representing $\rho$, $U$, $U^\dagger$ and $O$, respectively, as illustrated in Fig.~\ref{fig:mpo}.\\

\begin{figure}[ht]
    \centering
    \includegraphics[width=0.45\textwidth]{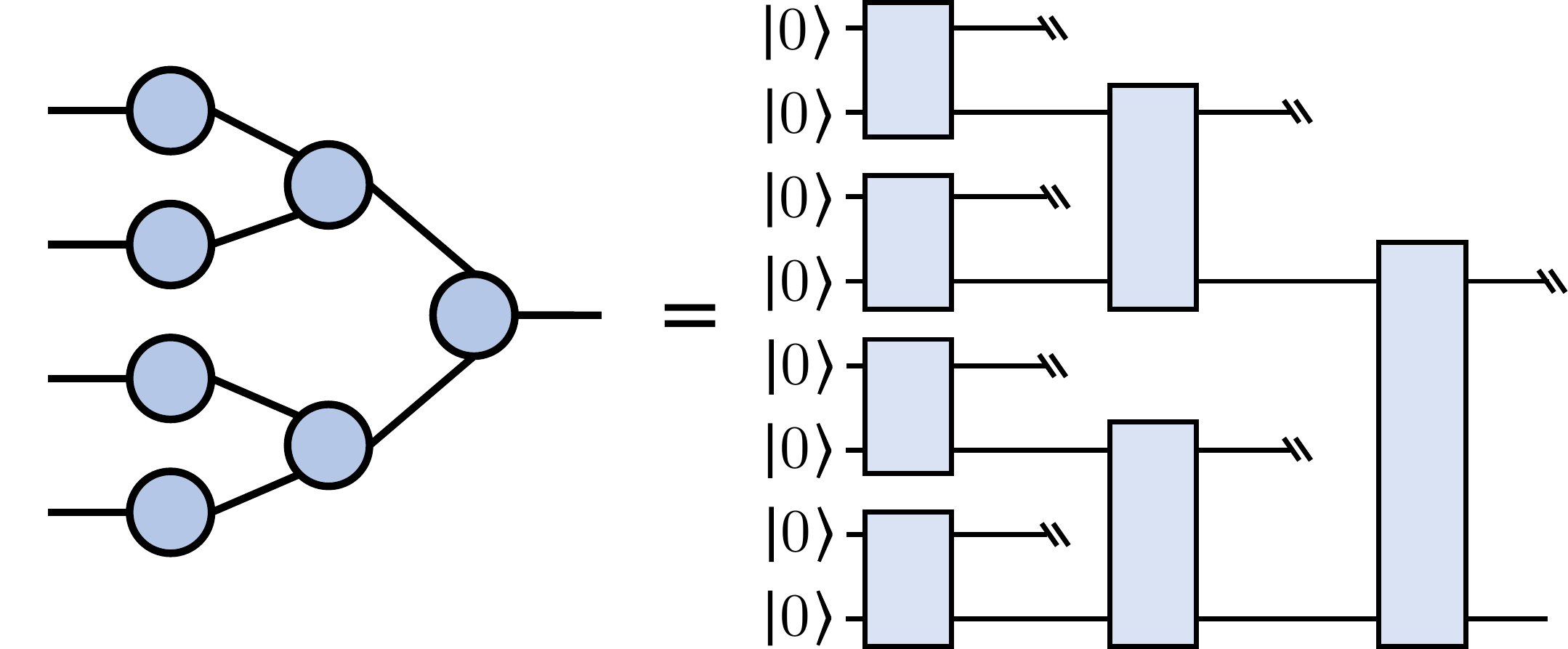}
    \caption{An example of the tree tensor network (TTN) architecture (\textit{left}) and the corresponding quantum circuit (\textit{right}). We use the symbol (\textbackslash\textbackslash) to clarify that a qubit is not affected by a particular unitary and is traced out.}
    \label{fig:TTN}
\end{figure}

Conversely, we can express certain tensor networks as quantum circuits, generating tensor-network quantum circuits. A tensor-network quantum circuit instructs a quantum computer to apply a transformation that is \emph{related} to its parent tensor network, inheriting the connectivity between the tensors in the network. Since quantum circuits must apply unitary operations, we only consider mapping the tensor network elements \emph{exactly} when the individual tensors are unitary operations. Otherwise, we allow the tensors to become general, undefined unitary operations. We refer to these resulting unitary operations as \emph{tensor blocks}. As for the parent tensor network's bond dimension $D$, this is captured by the number of bond qubits, $n_V$, shared by each tensor block, namely as $D=2^{n_V}$. As an illustration, Fig.~\ref{fig:MPS} shows a quantum circuit with an MPS architecture, and Fig.~\ref{fig:TTN} depicts a circuit following the structure of a tree tensor network (TTN).\\

A tensor-network quantum circuit is based on the shape and connectivity of its parent tensor network, for example, an MPS or TTN architecture, and not necessarily on specific tensor element values. Therefore, we can view a single tensor network architecture as a template for multiple possible circuits. We can obtain different circuits by changing the bond dimension of the parent tensor network or by varying the unitary operations corresponding to each tensor. Since each block can be an arbitrary unitary, it is crucial to define blocks that are compatible with (i) implementation on quantum hardware, (ii) fast simulation, and (iii) optimization strategies for quantum circuits. This is discussed in the following section.

\subsection{Meta-Ansatzes}

\begin{figure}[h]
    \centering
    \includegraphics[width=0.45\textwidth, trim=0.5cm 0cm 0cm 0.4cm, clip]{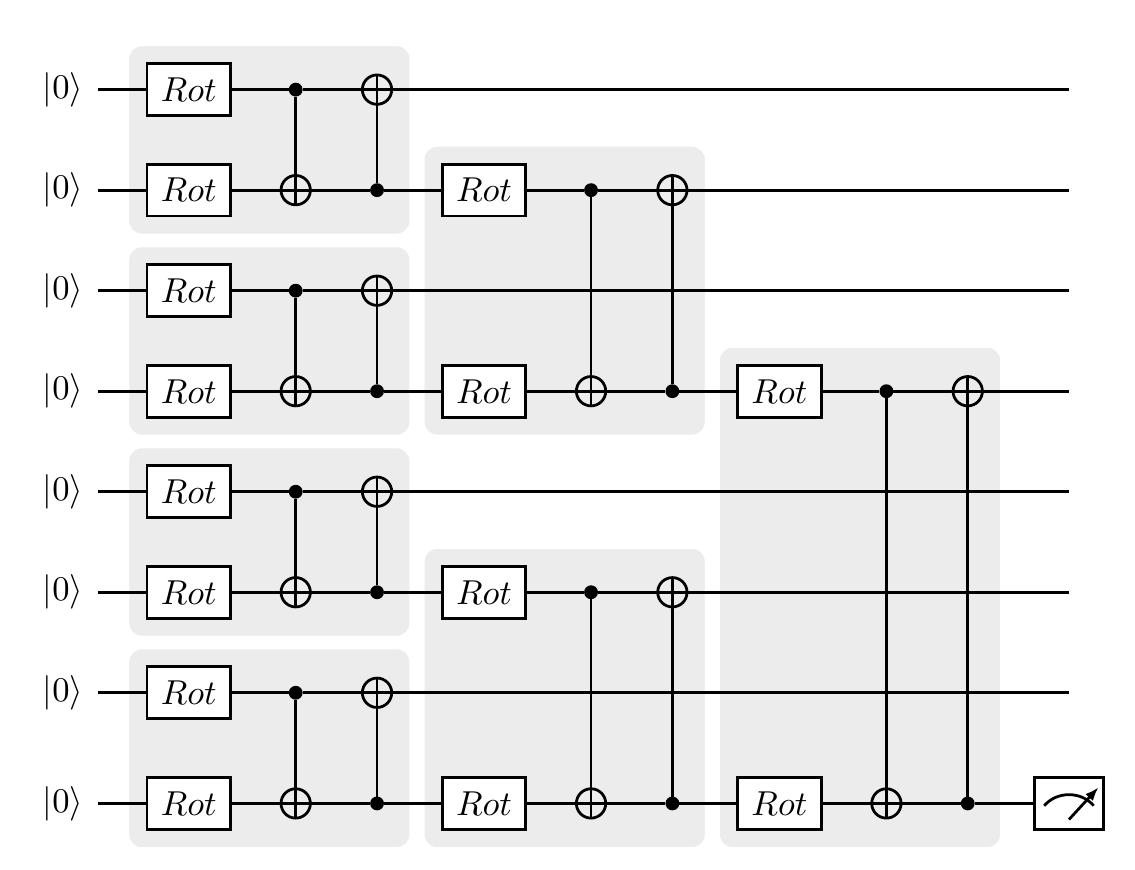}
    \caption{A tensor-network circuit as a meta-ansatz. We employ the same architecture as in Fig.~\ref{fig:TTN} and replace the tensor blocks with variational circuits consisting of single-qubit rotations and CNOT gates.}
    \label{fig:meta-TTN}
\end{figure}

One way to make tensor-network quantum circuits compatible with quantum hardware is to replace the unitary blocks with local circuits~\cite{haghshenas2022variational}. In this sense, just as a circuit ansatz is a strategy for arranging parametrized gates, tensor-network quantum circuits can be viewed as strategies for structuring smaller circuit ansatzes. They can therefore be interpreted as ansatzes of ansatzes, i.e., as \emph{meta-ansatzes}. This approach allows us to employ the same techniques used to design, implement, and optimize variational quantum circuits. This is illustrated in Fig.~\ref{fig:meta-TTN}, where we replace each block in a TTN circuit with a simple variational circuit. In summary, tensor networks can be used as a template to generate tensor-network meta-ansatzes. By replacing the tensor blocks in the meta-ansatz with parameterized circuits, we obtain a final variational quantum circuit. This circuit can then be simulated, implemented on hardware, and optimized as any other variational circuit.

\subsection{Designing Tensor-Network Quantum Circuits}

There are multiple ways to generate quantum circuits that relate to parent tensor networks~\cite{Huggins2019,cong2019quantum, haghshenas2021optimization, haghshenas2022variational, Ran_2020,Rudolph2022Synergy,Rudolph2022Decomposition}. We do not aim to rigorously reproduce the tensor network contraction with a quantum circuit, and instead endeavor to preserve certain features of the original tensor network: the number of tensors or operations, the connectivity between the operations, and the bond dimension of the connections. We now describe a procedure to generate tensor-network quantum circuits that maintain these features. 

\begin{figure}[t]
    \centering

    \includegraphics[width=0.45\textwidth]{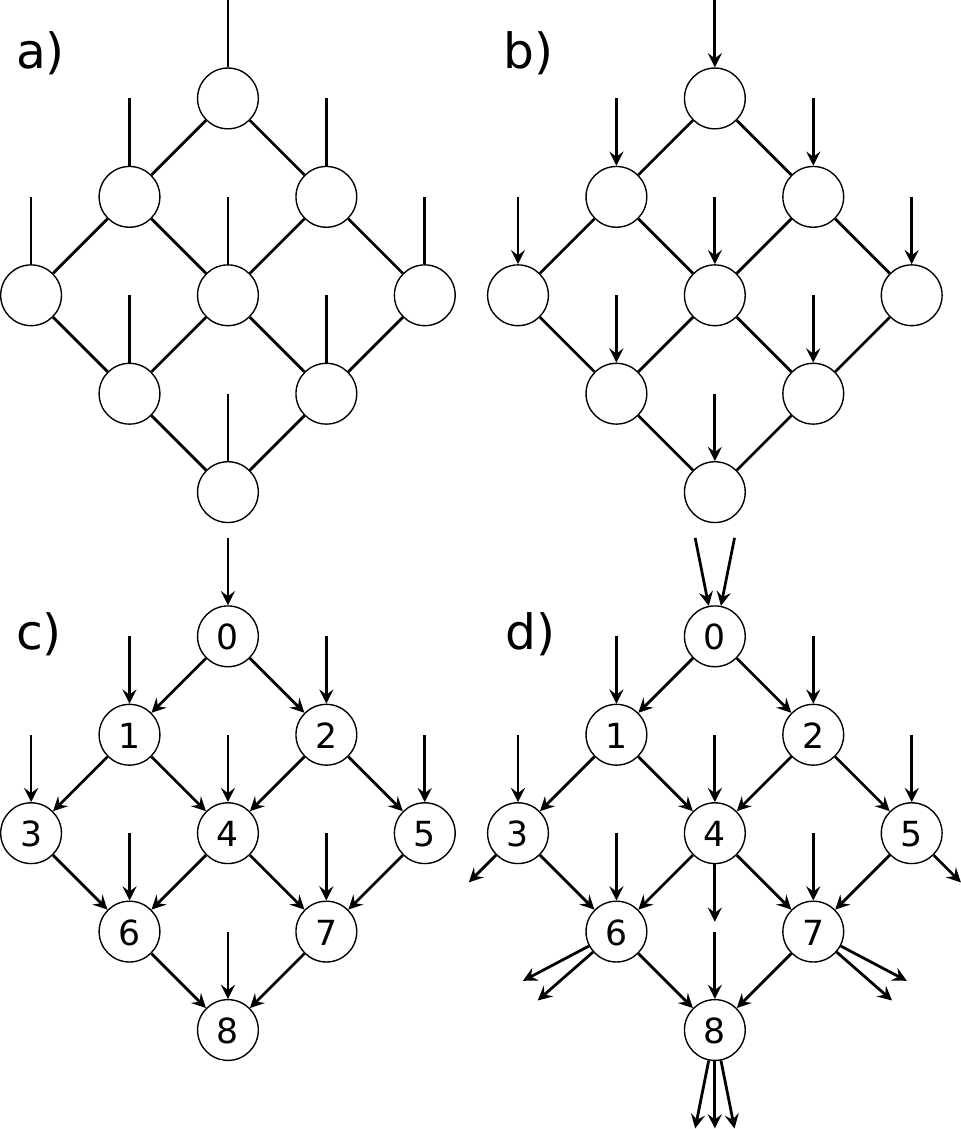}
    \caption{a) An example PEPS tensor network with bond dimension two, b) tensor-network quantum circuit design step 1: adding direction to the open edges in the tensor network diagram, c) steps 2 and 3: labeling the vertices and adding acyclic direction to the internal edges, d) step 4: adding edges to balance the in- and out-degree of each vertex.}
    \label{fig:tn>circuit_procedure}
\end{figure}

Consider a tensor network $T = (G, A)$ consisting of a collection of tensors $A = \{a(v) : v \in V \}$ and an undirected graph $G(E, V)$ defined by a set of edges $E$ and a set of vertices $V$, that admits open edges~\cite{peng2020simulating,robeva2019duality}. The vertices $V$ of $G$ represent tensors, whereas the tensor indices are depicted by the edges $E$ of $G$. For a tensor network $T = (G(E,V),A)$, we can construct a quantum circuit $C = G_c(E_c,V_c)$ such that $E\subseteq E_c$, $V=V_c$. A valid quantum circuit graph must be acyclic and have an equal number of incoming and outgoing open edges, representing the wires in the circuit. An index with dimension $d_e$ in the tensor network is replaced with $\lceil\log_2{d_e}\rceil$ qubit wires in the corresponding circuit. In other words, we can create a directed acyclic graph, as required for quantum circuits, while preserving the connections and number of vertices in the original tensor network.
Below we outline the full procedure in detail. 

\begin{enumerate}
    \item Choose the direction of the open edges in the tensor network. As in Ref.~\cite{Huggins2019}, this is informed by the desired application of the quantum circuit. For example, a generative machine learning task may use the open edges as outputs while a discriminative task may use them as inputs~\cite{Huggins2019}.
    \item Label the vertices with integers, $v$, in increasing order such that no two vertices have the same label.
    \item Set every edge $(v_i,v_j)$ to point from the lower integer to the higher integer vertex such that $v_i<v_j$. This ensures the graph is acyclic.
    \item Add new open edges to the circuit graph $G_c$: for each vertex where the number of incoming edges, $n_{in}$, is different from the number of outgoing edges, $n_{out}$, we add new directed edges such that $n_{in} = n_{out}$. This addresses the requirement that circuit operations must have equal numbers of incoming and outgoing wires. 
\end{enumerate}
The resulting graph represents a quantum circuit where the vertices are unitary operations and the edges are qubits.\\

As an example, Fig.~\ref{fig:tn>circuit_procedure} shows how we can obtain a quantum circuit graph from a projected entangled pair states (PEPS) tensor network graph. In Fig.~\ref{fig:tn>circuit_procedure}, the tensor network starts with nine tensors and nine open edges. This evolves into a quantum circuit with ten qubits and nine unitary operations. It is important to note that the number of qubits connecting two gates corresponds to the dimension of the index between the related tensors. While the example assumes a bond dimension of two for the parent tensor network, we can account for an increase in bond dimension by duplicating wires in the quantum circuit, such that $D=2^{n_V}$, as stated in Sec.~\ref{sec:tnqc}.

\subsection{Circuit Cutting}\label{sec:circuitcutting}
A quantum circuit can be executed on hardware or simulated classically, but hybrid methods also exist that trade off classical and quantum resources~\cite{bravyi2016trading, peng2020simulating, tang2021cutqc}. One of these techniques, circuit cutting~\cite{peng2020simulating,tang2021cutqc}, enables the execution of many-qubit circuits with few-qubit quantum devices, albeit at the expense of additional classical computation. The primary strategy in this technique is to divide large circuits into smaller fragments which are then evaluated on fewer-qubit devices. By evaluating these fragments over a large number of different configurations, we can obtain enough results to classically reconstruct the output of the original circuit.\\
\begin{figure*}
    \centering
    \includegraphics[width=0.85\textwidth]{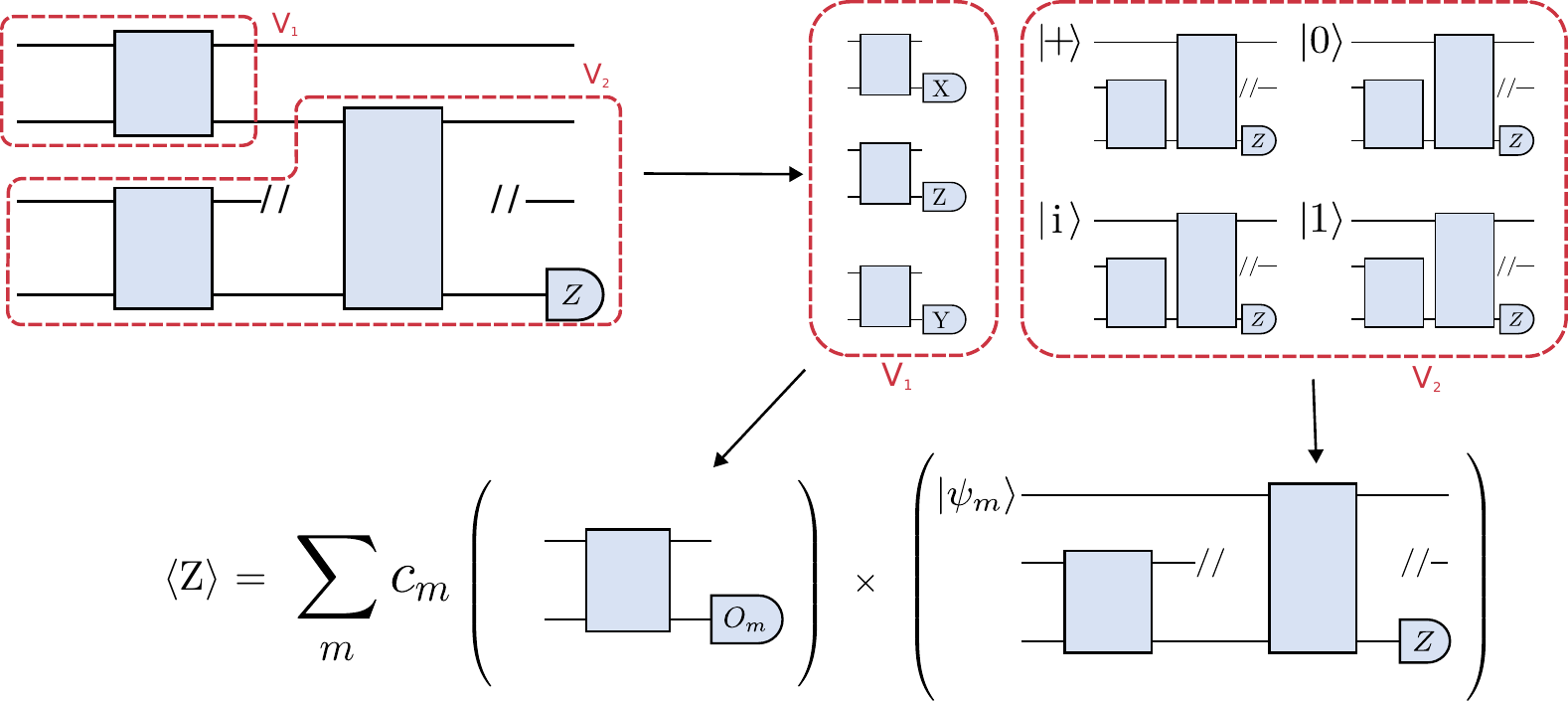}
    \caption{Circuit cutting and reconstruction procedure for a small TTN-shaped quantum circuit. \textit{Top}: the original circuit is partitioned into two fragments, $V_1$ and $V_2$. One fragment is executed with multiple different measurements, $\langle O_m\rangle$, while the other fragment is executed with multiple different initial states, $\ket{\psi_m}$. \textit{Bottom}: the results of the fragment executions are combined as dictated by Refs.~\cite{peng2020simulating,tang2021cutqc}. This summation is performed classical computer and returns the expectation value of a measurement on the original circuit.}
    \label{fig:ttn_partition}
\end{figure*}
More generally, recall that a quantum circuit can be described by a directed acyclic graph $G_c(V_c, E_c)$, where the nodes represent gates in the circuit and the edges represent wires. As we now explain, cutting a circuit is linked to partitioning this circuit graph. A partitioning $\Pi$ of a graph is a collection of subsets of vertices $V_{1}, V_{2}, \ldots, V_{k}\subset V_c$ such that every vertex in the graph is contained in exactly one subset. We refer to each subset $V_{i}$ as a graph fragment. The edges connecting different graph fragments correspond to wires that can be cut in the procedure, producing circuit fragments that can be executed separately. This is summarized in Fig.~\ref{fig:ttn_partition}.\\

The graph-based framework described above can be used to analyze the resource requirements of circuit cutting. Ref.~\cite{peng2020simulating} shows that the number of circuit executions needed to compute the expectation value of a tensor product of local observables of the form $O=\bigotimes_{i=1}^n O_i$ with precision $\varepsilon$ scales asymptotically as
\begin{equation}
    O \left(8^{3 d_{\rm max}} d_{\rm max}(k^{3} \log k) / \epsilon^{2} \right),
    \label{eq:circexecutionspeng}
\end{equation}
where $k$ is the number of fragments and $d_{\rm max}$ is the maximum number of edges between fragments. For the case of MPS circuits, this cost may be quadratically reduced following the techniques of Ref.~\cite{lowe2022fast}. \\

Looking into Eq.~\eqref{eq:circexecutionspeng}, we find that tensor-network quantum circuits are naturally suited for circuit cutting techniques: these circuits can be cut such that each tensor block results in a fragment and the exponent $d_{max}$ is kept fixed, allowing the circuit to be executed on few qubits while the number of circuits to evaluate scales polynomially with respect to the number of tensor blocks. For example, for architectures like MPS and TTN, the maximum number of edges between fragments $d_{\rm max}$ is equal to the number of bond qubits, $n_V$, that connect two adjacent blocks. Since $n_V$ is chosen in the design, it is possible to increase the number of qubits in the circuit while keeping $n_V$ constant. With a constant $n_V$, $k$ only increases linearly with respect to the total number of qubits. This means we can extend tensor-network quantum circuits as in Fig.~\ref{fig:ttn_partition} to more qubits and deeper circuits as in Fig.~\ref{fig:meta-TTN}, while the number of quantum circuit fragments we have to evaluate only increases polynomially with the number of qubits.

More precisely, consider an MPS circuit with bond dimension $2^{n_V}$, with blocks of $2n_V$ qubits, defined on $n$ total qubits and with a single Pauli $Z$ measurement on the bottom qubit. We can cut the circuit into its constituent blocks, meaning that we can evaluate the full circuit on a device with only $2n_V$ qubits. In this case, the number of different circuits that must be evaluated to reconstruct the measurement on the original circuit is given by:

\begin{equation}
    c_{\mathrm{MPS}} = 3^{n_V} + \left(\frac{n}{n_V}-3\right)4^{n_V}3^{n_V} + 4^{n_V}.
\end{equation}

For a TTN circuit with bond dimension $2^{n_V}$, with blocks of $2n_V$ qubits, defined on $n$ total qubits and with a single Pauli $Z$ measurement on the bottom qubit, this becomes:
\begin{equation}
    c_{\mathrm{TTN}} = \frac{3^{n_V}n}{2n_V} + 3^{n_V}(4^{2n_V})\left(\frac{n}{2n_V}-2\right) + 4^{2n_V}.
\end{equation}

\section{Numerical Demonstrations}\label{sec:numdemo}
To illustrate possible applications of tensor-network quantum circuits, we perform a series of numerical simulations. We combine circuit cutting techniques with tensor network circuits to classify synthetic data and then extend the model to image classification and object detection on industrial data. To carry out these experiments, we build on the open-source PennyLane library for quantum differentiable programming~\cite{PennyLane}.

\subsection{Circuit Cutting Simulation Times}

In this section, we start by benchmarking the runtime performance for the combination of circuit cutting techniques with tensor-network quantum circuits. We show how simulation time increases as we scale various tensor-network parameters.
\begin{figure}[h!]
    \centering
    \includegraphics[width=0.45\textwidth]{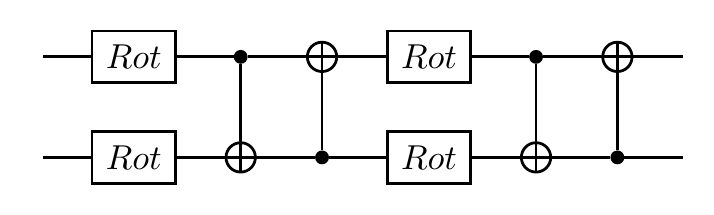}
    \caption{Two strongly entangling layers for two qubits. Each set of two rotation gates and two CNOT gates constitutes one strongly entangling layer. These layers can be repeated any number of times and extended to any number of qubits \cite{SEL}.}
    \label{fig:SEL}
\end{figure}

\begin{figure*}[t!]
    \centering
    \includegraphics[width=0.9\textwidth]{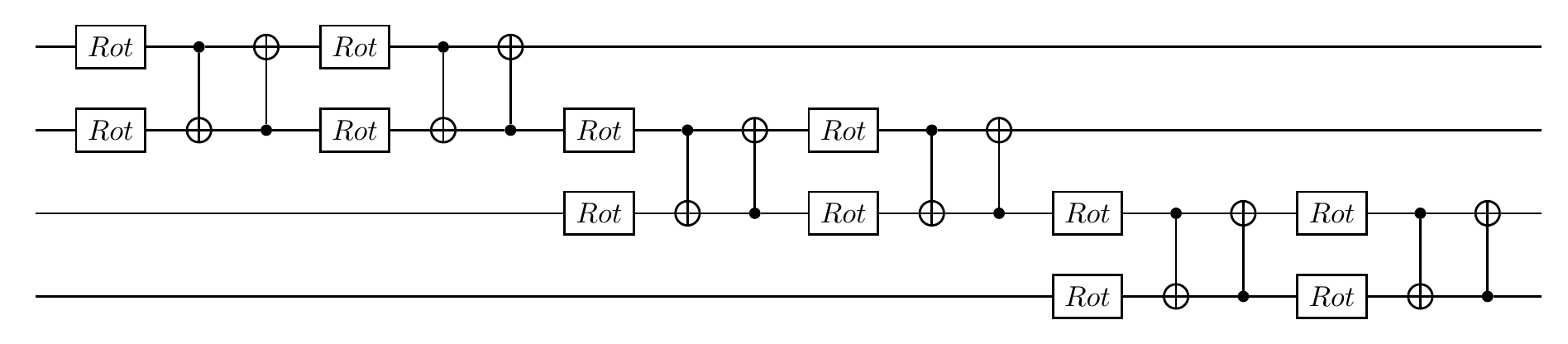}
    \caption{MPS meta-ansatz with two strongly entangling layers replacing each unitary tensor block.}
    \label{fig:MPS-SEL}
\end{figure*}
More specifically, we design an MPS quantum circuit as in Fig.~\ref{fig:MPS}, where the unitary blocks are replaced with two strongly entangling layers \cite{SEL} like the ones in Fig.~\ref{fig:SEL}. An example of a resulting MPS with the unitary blocks specified as strongly entangling layers is given in Fig.~\ref{fig:MPS-SEL} for four qubits. We then add a Pauli $Z$ measurement on the bottom qubit and simulate the resulting circuit. PennyLane's \verb|MPS| template can be used to produce an MPS circuit with user-defined circuit blocks, number of bond qubits, and total number of qubits. By defining a block that includes the strongly entangling layers template, we can define a circuit like in Fig.~\ref{fig:MPS-SEL}. We then use PennyLane's circuit cutting functionality to separate the circuit into its individual tensor blocks, add the required state preparations and observables, evaluate them, and reconstruct the original circuit result. This is done automatically when the \verb|cutcircuit| decorator is applied to a PennyLane circuit. The simulations are performed for various configurations of the bond dimension, block size, and the total number of qubits. For an example of how to use PennyLane to simulate circuit cutting, see the Appendix.\\

\begin{figure*}[]
    \includegraphics[width=0.3\textwidth]{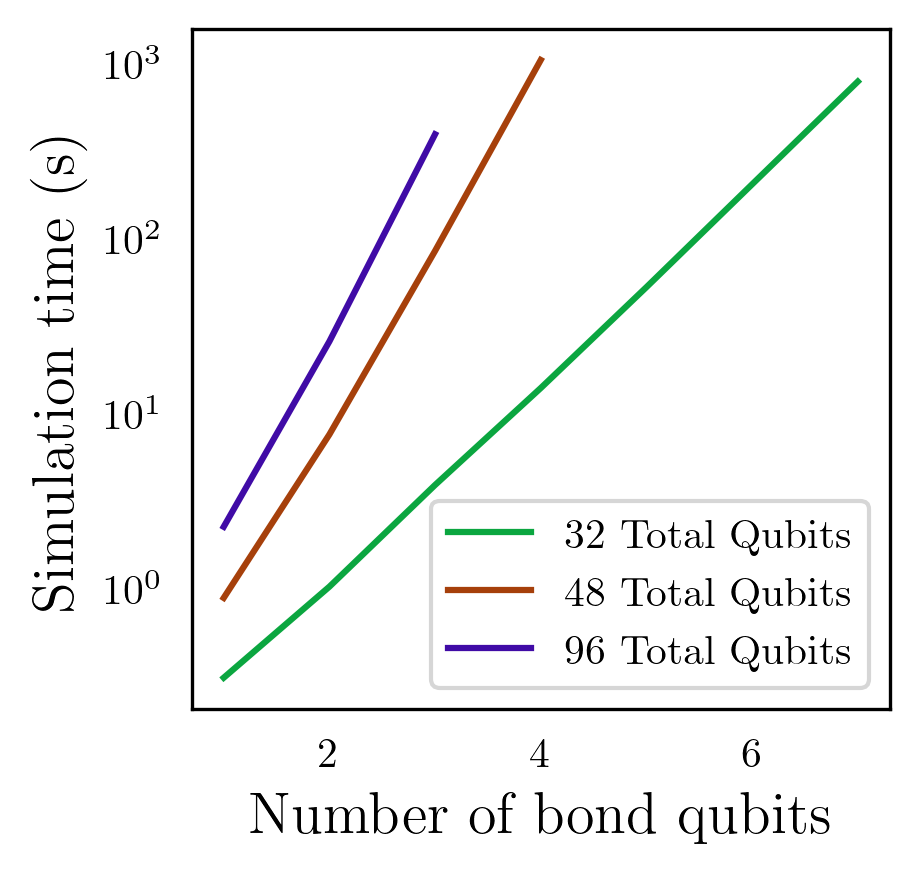}
    \includegraphics[width=0.3\textwidth]{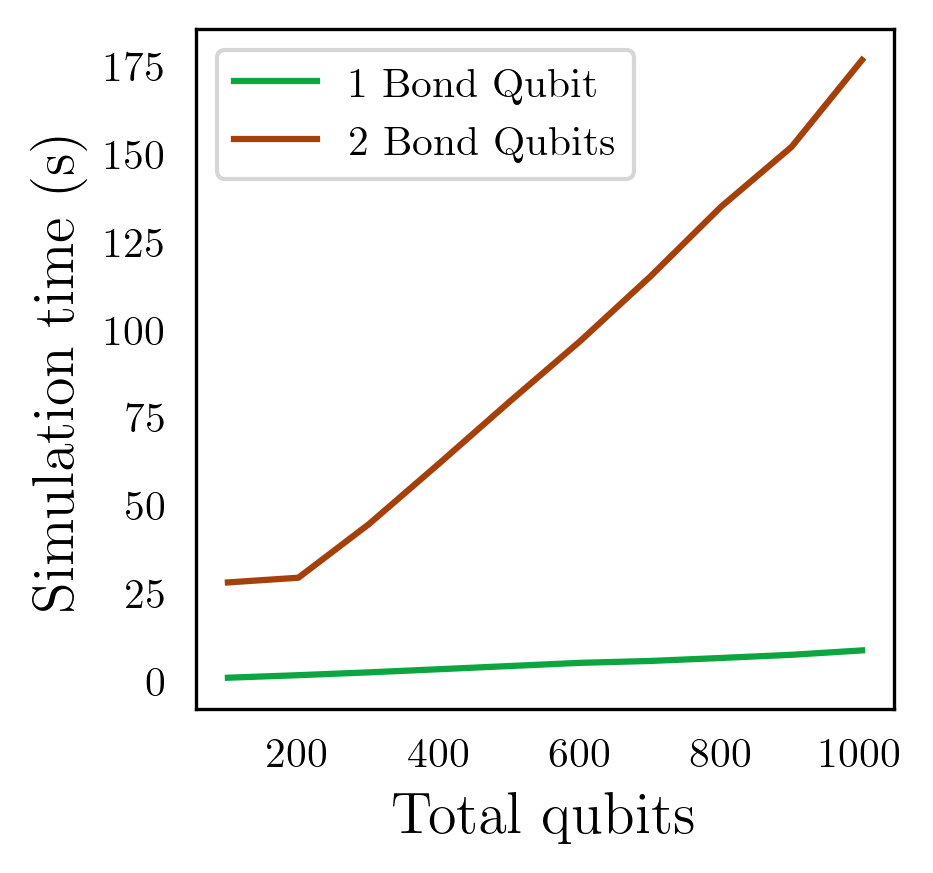}
    \includegraphics[width=0.3\textwidth]{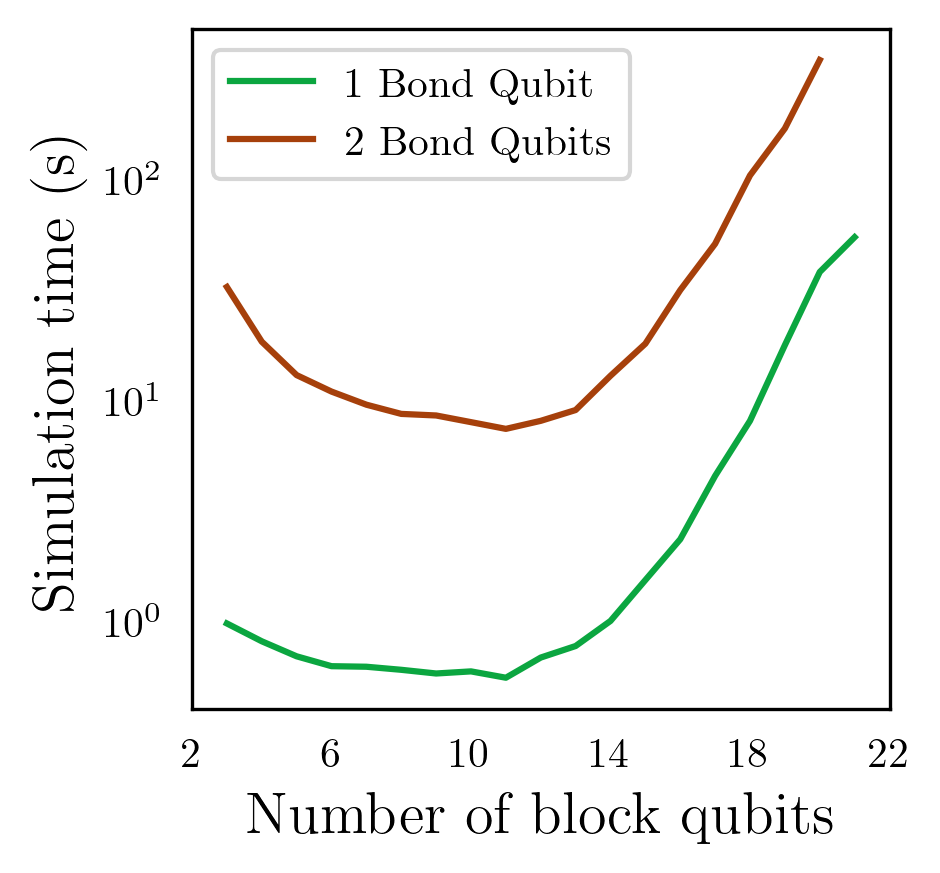}
    \caption{\textit{Left}: The simulation time of an MPS circuit increases exponentially with the number of bond qubits, regardless of the total number of qubits. For this data, we used 16-qubit blocks. Note that the total number of qubits can vary slightly as the number of bond qubits changes. The MPS shape dictates the variation in qubit numbers, e.g., an MPS circuit with 16-qubit blocks and two bond qubits per block can only result in circuits with $16+14n$ qubits, where $n$ is a positive integer. \textit{Middle}: At a constant number of bond qubits and five block qubits, the simulation time increases linearly with the total number of qubits. \textit{Right}: For a constant total circuit size of 100 qubits, increasing the size of the tensor blocks initially reduces the simulation time and then increases it. This is an artifact of how the tensor blocks are defined. Initially, increasing the number of block qubits reduces the total number of circuits to simulate during circuit cutting. However, as the size of the blocks increases, the time gained by having larger circuits surpasses the time saved by having fewer circuits. All simulations are performed on a personal laptop computer with 16 GB of RAM and a four-core i7-1185G7 processor operating at 3.00GHz.}
    \label{fig:simulationTimes}
\end{figure*}

Following the equations in Sec.~\ref{sec:circuitcutting}, we find that the simulation time increases polynomially with the total number of qubits and exponentially with the number of bond qubits. This is shown in Fig.~\ref{fig:simulationTimes}. Overall, applying circuit cutting to an MPS quantum circuit enables the simulation of a large number of qubits as long as the number of bond qubits is kept low. For the simple structure of these circuits, this performance could also be achieved with simulations based on classical tensor network techniques~\cite{markov2008simulating}, but circuit cutting provides a path toward executing large circuits using small quantum computers.

\subsection{Bars and Stripes}

Here we demonstrate how to use a tensor-network quantum circuit to perform image classification tasks. The problems we study are well-known and can be routinely solved with classical methods. Our purpose is not to compete with such techniques, but rather to guide readers on example applications of tensor-network quantum circuits.

\begin{figure}
    \centering
    \includegraphics[width=0.45\textwidth]{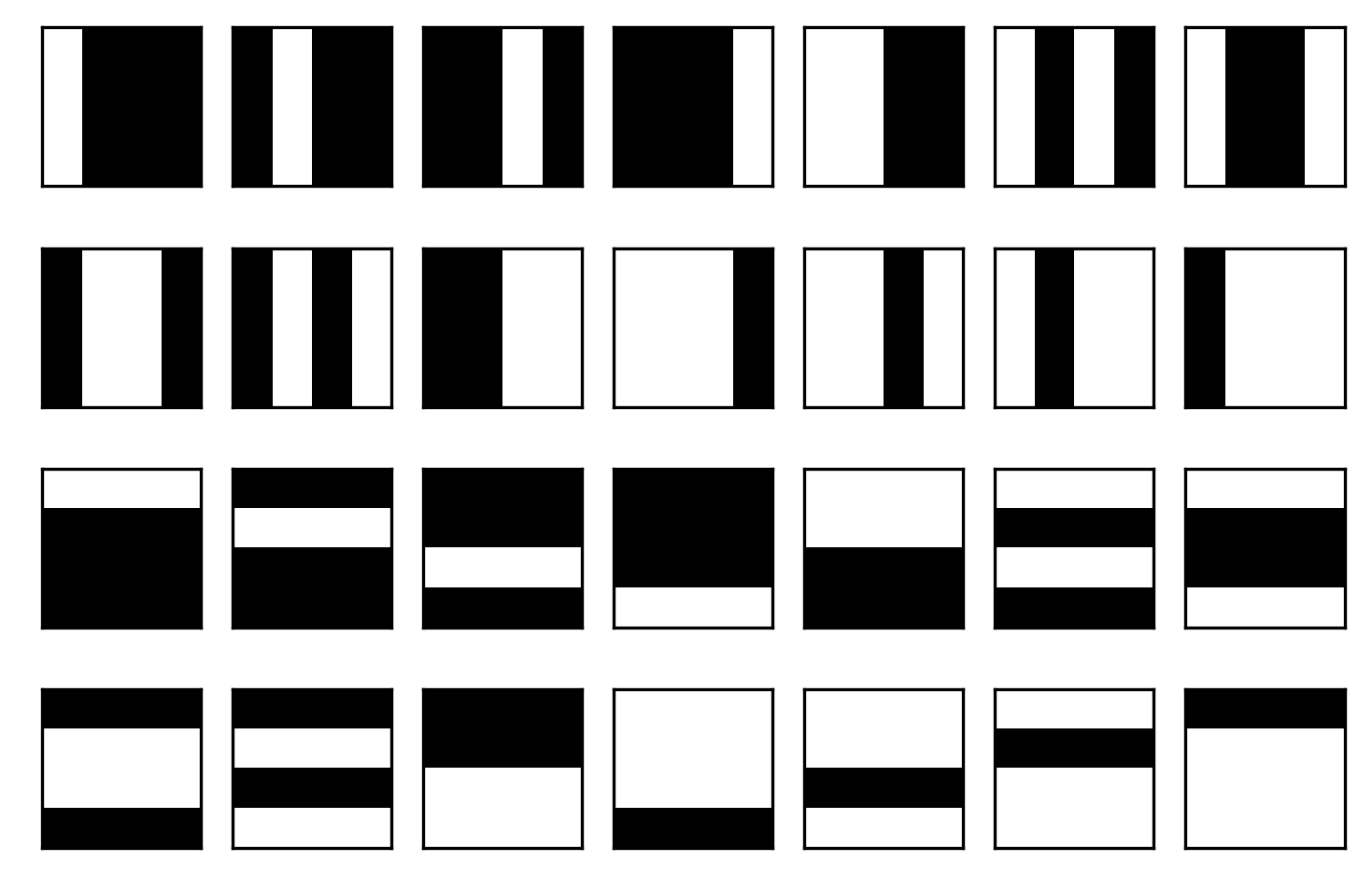}
    \caption{Bars and stripes data set for $4\times4$ pixel images.}
    \label{fig:BAS_Dataset}
\end{figure}

The bars and stripes data set is an example of synthetic data often used to develop proof-of-principle machine learning algorithms. As shown in Fig.~\ref{fig:BAS_Dataset}, a bars and stripes instance is composed of binary black and white images of size $n \times n$ pixels, where either all pixels in a column have the same color (bars) or all pixels in a row have the same color (stripes)~\cite{BAS}. The classification task is to output the correct label, bars or stripes, for any input image from the data set. To perform this task, we implement a quantum circuit consisting of an encoding operation to input the image, a parameterized tensor-network quantum circuit to process it, and a measurement to obtain the label. Since many design choices are required to implement this framework, we summarize these in the following list: 
\begin{enumerate}
    \item We choose the amplitude encoding \cite{schuld2021machine} to encode the normalized datapoint $x$ of pixel information into the amplitude of a $n$-qubit quantum state
    \begin{align*}
        \ket{\psi_x} = \sum^{2^n}_{i=1}x_i\ket{i},
    \end{align*}
    with $x_i$ the $i$-th element of $x$ and $\ket{i}$ the $i$-th computational basis state.
    \item We choose a tree tensor network architecture because its hierarchical structure is suited to perform image processing tasks like convolution and pooling~\cite{Huggins2019,TTNvsMPS_MLUnitary,TTNvsMPS_Expressive}.
    \item We use two strongly entangling layers to replace the unitary blocks in the circuit because they are expressive~\cite{SEL} and experimentation showed that two layers can reach 100\% classification accuracy for this application.
    \item We limit the individual blocks to two qubits, to reduce computation time while still reaching 100\% training accuracy.
    \item We make a Pauli $Z$ measurement on the bottom qubit to obtain the labels. When an input image results in a Pauli $Z$ measurement of positive one, we label that image ``bars'' and when the Pauli $Z$ measurement is negative one, we label the image ``stripes''. For multiple measurements, we use the expectation value, such that when $\langle\sigma^Z\rangle > 0$, we label the image ``bars'' and when $\langle\sigma^Z\rangle < 0$, we label the image ``stripes''. In other words, we choose the most-frequently-sampled label. 
    \item We choose 14 training images and 14 test images from the bars and stripes data set.
    \item We use the loss function \begin{equation} \label{eq:lossfunction}
    loss=\sum_i{(1+10e^{7p_i})^{-1}},
\end{equation}
    where the index $i$ iterates over the images in the data set, and $p_i$ is the probability of obtaining the correct label when sampling the circuit with image $i$ as input. This loss function favors a good probability of sampling correct labels over many images rather than a very high probability over a few images. The parameter $p_i$ can be calculated from the Pauli $Z$ expectation value as: 
    \begin{equation}\label{eq:pi_expval}
    p_i = 1- \left |{\frac{1 - \bra{\phi_i} \sigma^z_n\ket{\phi_i}}{2}-\ell} \right|,
\end{equation}
    Where $\sigma^z_n$ is the Pauli $Z$ operator applied to the $n$-th qubit, $\ket{\phi_i}$ is the final state of the qubits after running the circuit for image $i$, and $\ell=0,1$ is the correct image label, taking a value of zero for bars and one for stripes.
    \item We use the Simultaneous Perturbation Stochastic Approximation (SPSA) algorithm to optimize the circuit, with hyperparameters $\alpha=0.602$ and $\gamma=0.101$~\cite{SPSA}.
    \item We use PennyLane~\cite{PennyLane} templates to design the tensor-network circuits and Jet~\cite{Jet} to simulate the circuits using optimized task-based tensor contraction.
\end{enumerate}

\begin{figure}[t!]
    \centering
    \includegraphics[width=0.45\textwidth, trim=0.4cm 0cm 0cm 0cm, clip]{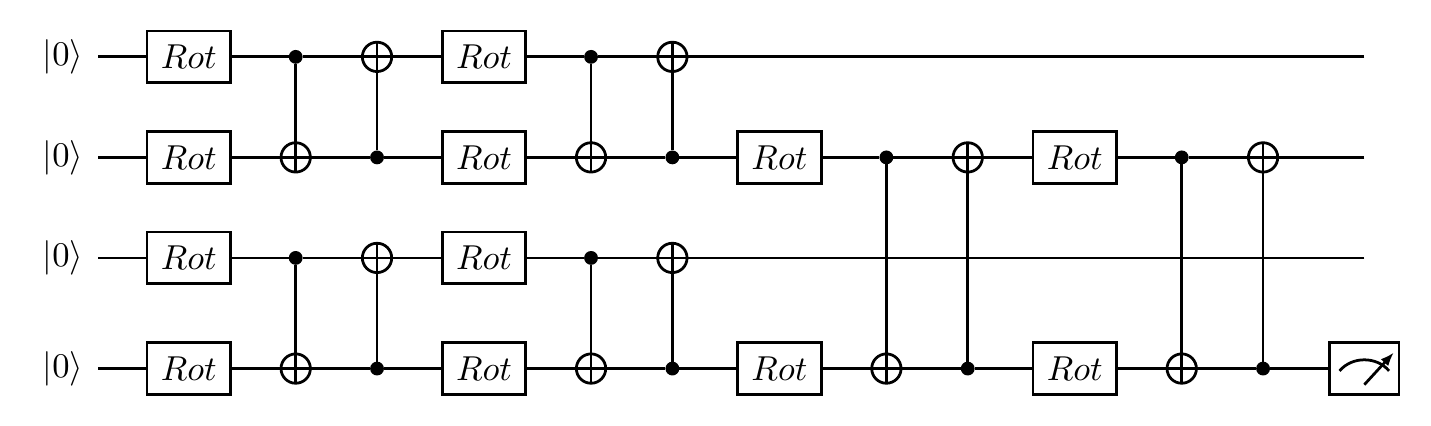}
    \caption{A tree tensor quantum circuit with $n_V=1$ and two entangling layers applied to four qubits. Single-qubit gates apply arbitrary Bloch rotations of user-defined value $\omega$ in the $Z$ axis, $\theta$ in the $Y$ axis, and $\phi$ in the $X$ axis. These rotations can be optimized such that the circuit classifies $16\times 16$ pixel images. The circuit can also be extended to more qubits, enabling the classification of larger images.}
    \label{fig:fullTTNcircuit}
\end{figure}

An example circuit on four qubits following these design choices is shown in Fig.~\ref{fig:fullTTNcircuit}.\\

\begin{figure}[h]
    \centering
    \includegraphics[width=0.45\textwidth]{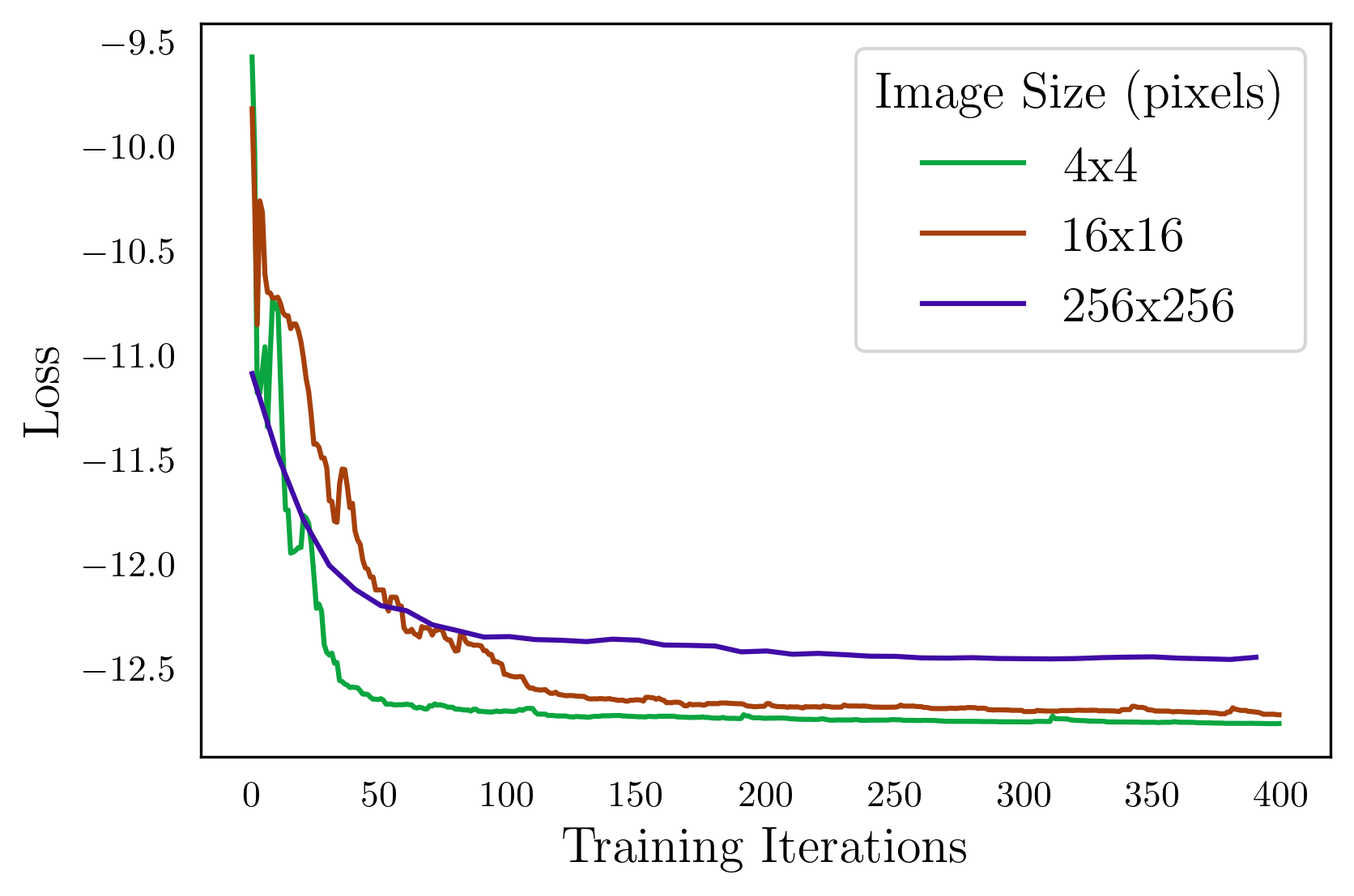}
    \includegraphics[width=0.45\textwidth]{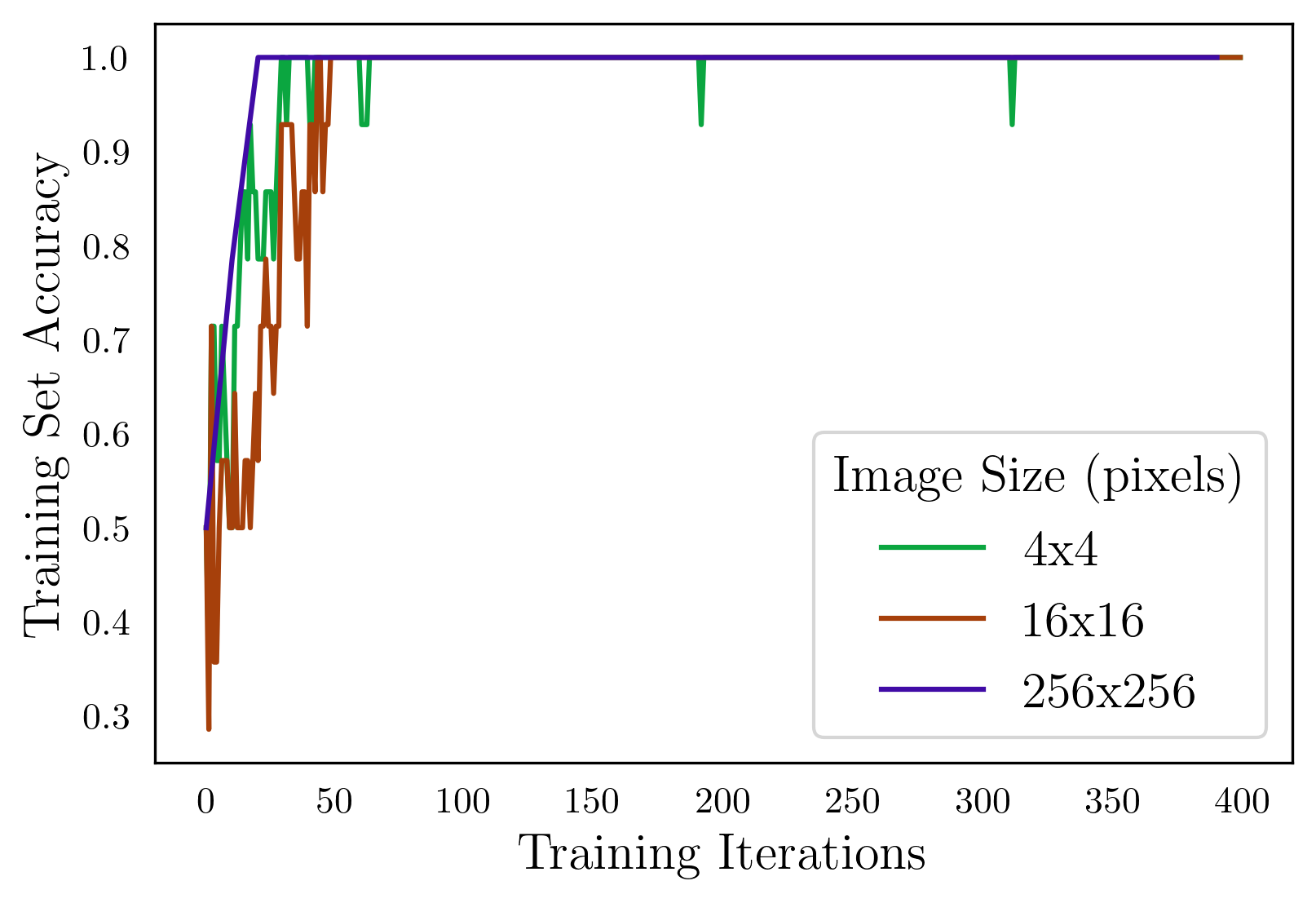}
    \includegraphics[width=0.45\textwidth]{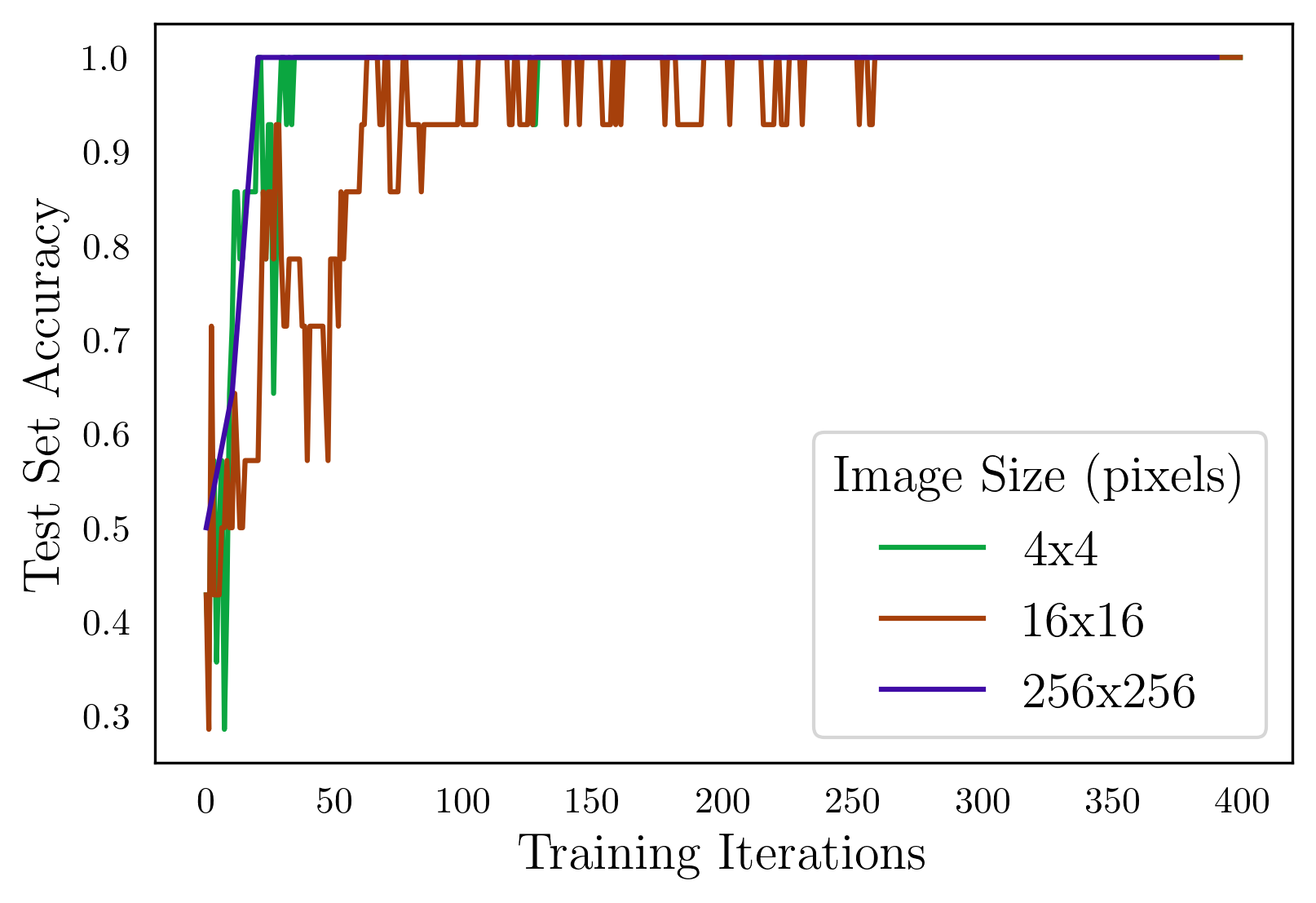}
    \caption{Evolution of loss, training accuracy, and test accuracy while training a tree tensor quantum circuit on various image sizes. The quantum circuit reaches 100\% accuracy for all sizes within 400 training iterations.}
    \label{fig:BAS}
\end{figure}

Next, we train the tree tensor network circuit on images of size $4 \times 4$ pixels, then extend to $16\times 16$ pixels, and finally reach $256 \times 256$ pixels. Under these design conditions, we can train the circuit to reach 100\% classification accuracy for both the training and test sets. This is most likely due to the simplicity of the task, as we will see in the next section. These results are summarized in Fig.~\ref{fig:BAS}.

\subsection{Welding Defects}\label{sec:weldingdefects}
\begin{figure}[h!]
    \centering
    \frame{\includegraphics[width=0.23\textwidth, trim=0.5cm 1cm 2cm 0.4cm, clip]{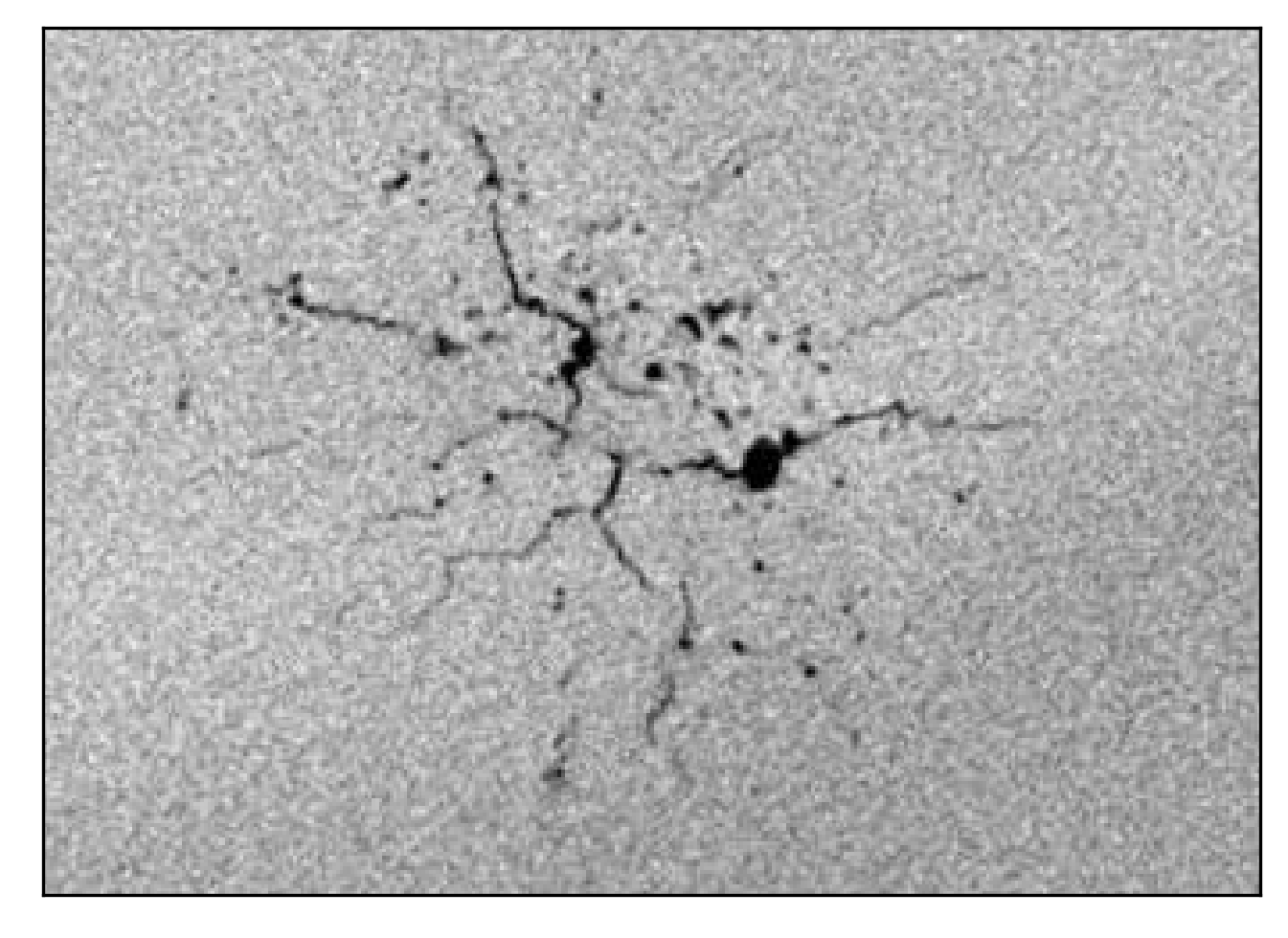}}
    \frame{\includegraphics[width=0.23\textwidth, trim=0.5cm 1cm 2cm 0.4cm, clip]{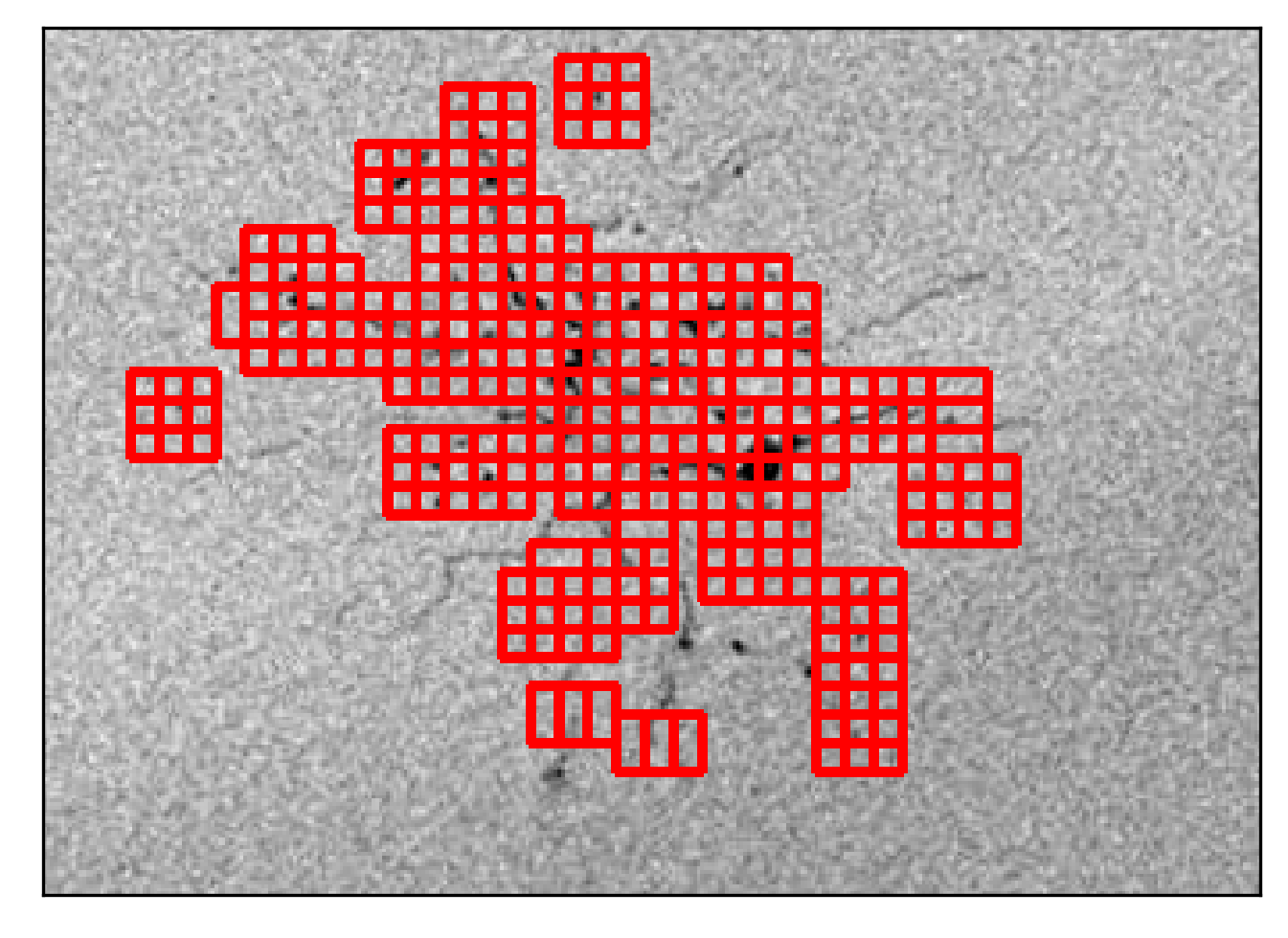}}
    \frame{\includegraphics[width=0.23\textwidth, trim=0.5cm 1cm 2cm 0.4cm, clip]{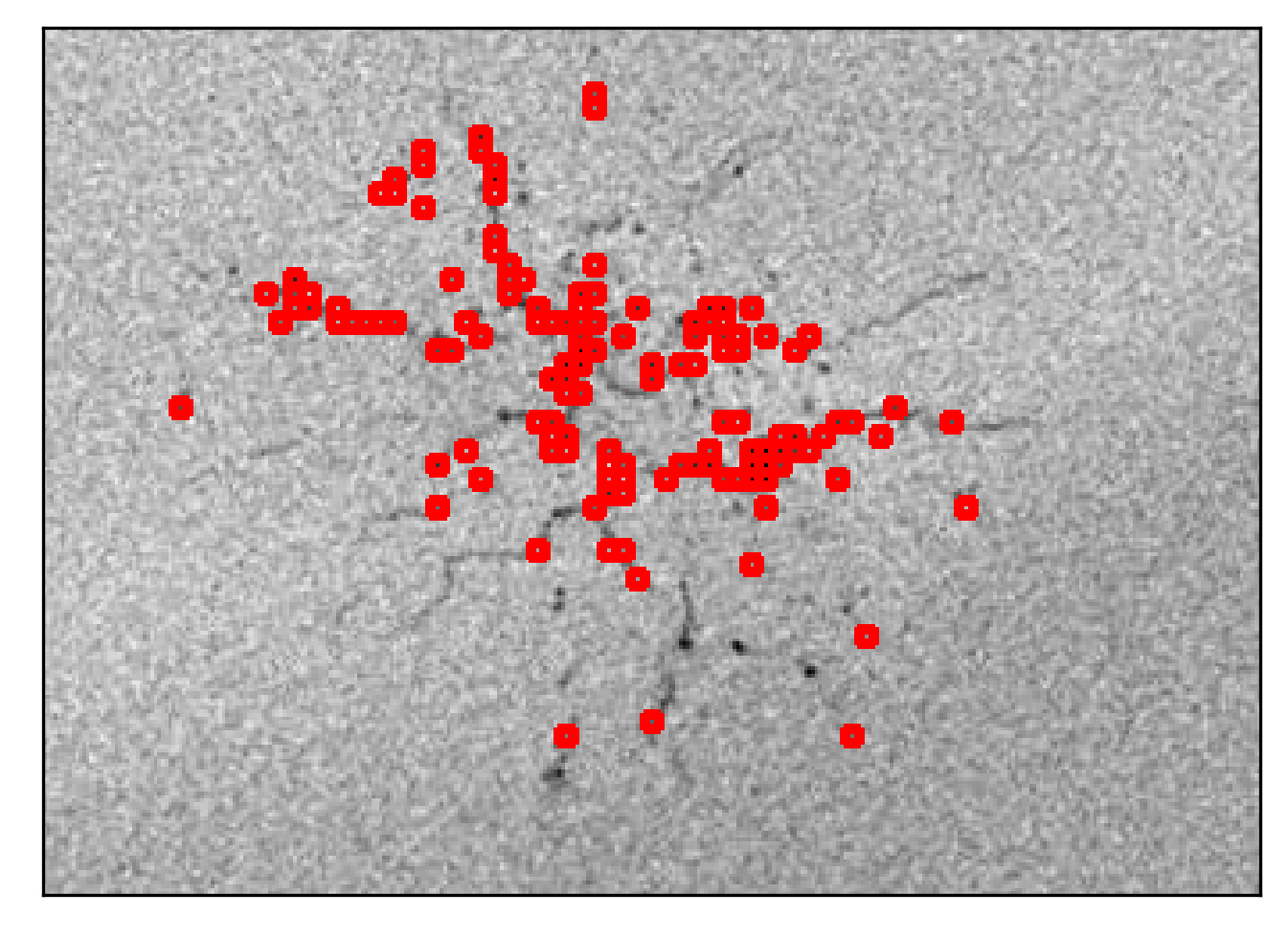}}
    \frame{\includegraphics[width=0.23\textwidth, trim=0.5cm 1cm 2cm 0.4cm, clip]{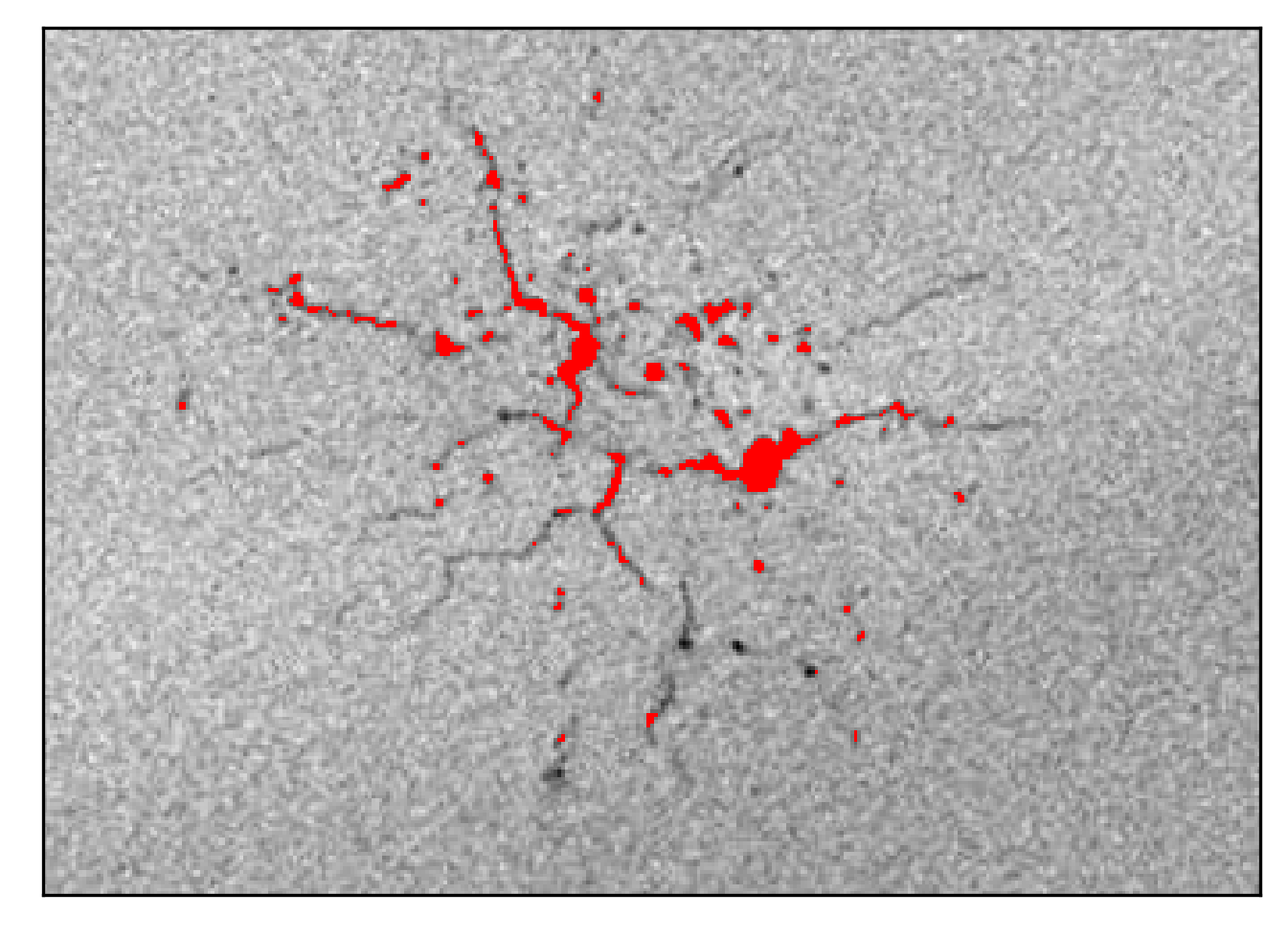}}
    \caption{Welding defects example image, followed by coarse-grained bounding boxes, fine-grained bounding boxes, and highlighted defects. This figure illustrates the full defect detection procedure and the results at each step. The first image is identified as having defects by a $256 \times 256$ pixel classifier. This classifier is trained on manually chosen images, obtaining 79\% accuracy on a training set and 71\% accuracy on a test set. After segmenting the original image (top left) into $16\times16$ pixel images, classifying each segment, and highlighting the segments with the defect label, we obtain the top right image. This classifier is trained on manually chosen images, obtaining 100\% accuracy on the training set and 79\% accuracy on the test set. Further segmenting these highlighted sections into $4\times4$ pixel images and classifying yields the bottom left image. This classifier reaches 100\% accuracy on a similarly defined training set and 71\% accuracy on a test set. To visualize the percentage of the defect that is captured, we classically post-process the highlighted sections, converting black pixels in these areas to red and showing the result in the bottom right image. While the algorithm occasionally misses small defects, the more severe portions are identified.
    }
    \label{fig:WeldExample}
\end{figure}

In this section, we extend the previous image classifier circuit to perform object detection on weld images toward implementing quality control systems. Welding is a standard method to fuse two portions of metallic material. It consists of partially melting the metal to attach the materials and allowing it to solidify. During this process, defects can weaken the connection between the materials. The welding defects data set contains cross-sectional X-ray images of the fused portion in different welded structures. The flaws in the images appear as very dark or black cracks and bubbles, as seen in Fig.~\ref{fig:WeldExample}. The goal here is to determine the severity and extent of the defects by quantifying the area they occupy using a tree tensor network circuit that identifies the size and location of the defect.\\

This object detection is performed in three steps. We first classify an image as containing defects or not. We then use the sliding windows method~\cite{dalal2005histograms} to segment the image into many smaller pictures and classify each segment with $16\times 16$ pixel segments to propose defect areas. In the final step, we re-segment the proposed areas into $4 \times 4$ pixel images to identify the individual defects in the proposed areas. This strategy requires training three image classifiers, one for each step. The overall defects-versus-no-defects classifier is a 16-qubit tree tensor network circuit. To train this classifier, we manually select 14 defect and 14 non-defect images and divide them into a training set and test set. We crop and resize the images to $256\times 256$ pixels before inputting them to the circuit. This is done both to fit the 16-qubit size requirement and because the defects are typically in the center of the images. \\

For the second classifier, we use an eight-qubit circuit that can process the $16\times16$ pixel segments. To train this circuit, we segment an image with defects into $16\times 16$ pixel images and manually select 14 segments with defects and 14 without defects. Finally, we repeat the previous procedure with a four-qubit circuit and $4 \times 4$ pixel images. Due to the smaller size, the four-qubit and eight-qubit circuits are significantly faster to train and simulate than the 16-qubit circuit.\\

In summary, the complete defect detection strategy involves first running a 16-qubit tree tensor circuit to classify whether the center $256 \times 256$ pixel portion of the weld image has a flaw, running the sliding window algorithm to classify $16\times 16$ pixel segments of the image to propose sections with flaws, and finally running the sliding window algorithm to classify $4\times 4$ pixel segments of the proposed area. Once the final $4\times 4$ pixel segments are classified, we can classically select the black pixels in those segments and convert them to red to highlight the detected defect area. The results of running the entire algorithm on an example image are shown in Fig.~\ref{fig:WeldExample}. This figure demonstrates that the procedure indeed detects the defect areas.

\section{Conclusion}\label{sec:conclusion}
In this work, we have provided an overview of how to apply tensor-network architectures to the design of variational quantum circuits. We implement these variational circuits to address illustrative industry-relevant problems. The results serve as examples of potential proof-of-principle use cases for existing quantum hardware and simulators. Additionally, the results show how combining circuit cutting with tensor-network quantum circuits can improve the scale of quantum systems that can be simulated in classical computers. Additionally,  the results can be leveraged to execute large tensor-network quantum circuits on small quantum devices. Moreover, we find that simple image-classification tasks can be performed on quantum computers via this method.\\

Additional work must be done comparing the performance of tensor network quantum circuits to classical alternatives. While we do not anticipate that tensor-network quantum circuits will outperform classical algorithms for the investigated image-processing applications, the tensor-network quantum circuit framework may help study the relationships between data structure and the design of quantum algorithms. \\



\section*{ACKNOWLEDGEMENT}
We thank Mikhail Andrenkov, Burak Mete, Sepehr Taghavi, and Trevor Vincent, for fruitful discussions. CAR and JK are partly funded by the German Ministry for Education and Research (BMB+F) in the project QAI2-Q-KIS under grant 13N15583.

\bibliographystyle{apsrev}
\bibliography{references.bib}

\appendix
\section{Code}\label{code:circuit_cutting}

\begin{lstlisting}[language=Python, caption=An example of generating MPS meta-ansatz circuit with PennyLane]
import pennylane as qml
from pennylane import numpy as np

np.random.seed(1)

def block(weights, wires):
    qml.StronglyEntanglingLayers(weights, 
    wires)
    qml.WireCut(wires=wires)

shape = qml.StronglyEntanglingLayers.shape(
    n_layers=2,n_wires=2)
    
template_weights=[
    np.random.random(size=shape)]\
    *qml.MPS.get_n_blocks(wires=range(4),
    n_block_wires=2)

dev = qml.device('default.qubit',wires=2)
@qml.cut_circuit
@qml.qnode(dev)
def circuit():
    qml.MPS(wires=range(4),
        n_block_wires=2,block=block,
        n_params_block=3,
        template_weights=template_weights)
    return qml.expval(qml.PauliZ(wires=3))
\end{lstlisting}
\end{document}